\documentclass[aps,preprint,nofootinbib,tightenlines,superscriptaddress]{revtex4}
\pdfoutput=1
\usepackage{epsfig}
\usepackage{amsmath}
\usepackage{graphicx}
\newcommand{\beq}{\begin{equation}}
\newcommand{\eeq}{\end{equation}}
\newcommand{\bqa}{\begin{eqnarray}}
\newcommand{\eqa}{\end{eqnarray}}
\newcommand{\nnb}{\nonumber\\}

\begin{document}

\title{Factorization for radiative  heavy quarkonium decays  into scalar Glueball }

\author{Ruilin Zhu~\footnote{Email:rlzhu@sjtu.edu.cn.}}
\affiliation{
 INPAC, Shanghai Key Laboratory for Particle Physics and Cosmology, Department of Physics and Astronomy, 
 Shanghai Jiao Tong University, Shanghai 200240,   China}
 \affiliation{
State Key Laboratory of Theoretical Physics, Institute of Theoretical Physics, Chinese Academy of Sciences, Beijing 100190, China}
\affiliation{
CAS Center for Excellence in Particle Physics, Institute of High Energy Physics, Chinese Academy of Sciences, Beijing 100049, China}
\begin{abstract}
We establish the factorization formula for scalar Glueball production through radiative decays of  vector states of heavy quarkonia,
e.g. $J/\psi$, $\psi(2S)$ and $\Upsilon(nS)$, where the Glueball mass is much less  than the parent heavy quarkonium mass.
The factorization is demonstrated explicitly at one-loop level through the next-to-leading order (NLO) corrections to the hard kernel,
the non-relativistic QCD (NRQCD) long-distance matrix elements (LDMEs) of the heavy quarkonium, and the
light-cone distribution amplitude (LCDA) of scalar Glueball. The factorization provides a comprehensive theoretical
approach to investigate
Glueball production in the radiative decays of vector states of heavy quarkonia and determine the physic nature of Glueball.
We  discuss the scale evolution equation of LCDA for scalar Glueball. In the end, we extract the value of the decay
constant of Scalar Glueball from Lattice QCD calculation
and analyze the mixing effect among  $f_0(1370)$, $f_0(1500)$ and $f_0(1710)$.

\end{abstract}

\maketitle

\section{Introduction}

Glueball, as a kind of color-confined state with two or more than two gluons, is one of the most important
expectation in Quantum Chromodynamics (QCD). The non-Abelian interactions among gluons tell us the
existence of Glueball and there is no hidden principle to forbid this kind of state up to now. Many theoretical
approaches have been
employed to investigate one of the most important quantum numbers of Glueball, i.e. its mass, e.g. Lattice QCD~\cite{Albanese:1987ds,Morningstar:1999rf,Chen:2005mg,Gregory:2012hu},
QCD Sum Rules~\cite{Bagan:1990sy,Liu:1993it,Huang:1998wj,Qiao:2014vva},
Supergravity Dual~\cite{Csaki:1998qr}, Top-down Holographic Dual~\cite{Brunner:2015oqa}, Rotating Closed Strings~\cite{Sonnenschein:2015zaa}, MIT Bag Model~\cite{Carlson:1982er}. In these approaches, the scalar Glueball
is expected generally
to populate the low enery region from 1{GeV} to 2.2{GeV}, which is also a region rich in $q\bar{q}$ states.
 A famous criterion to distinguish Glueball from the conventional $q\bar{q}$ states is that the width of Glueball is narrow
  from the large-$N_c$ argument~\cite{Cheng:2015iaa,He:2015owa,Wang:2009rc}, where the decay width of the $q\bar{q}$ states is
   proportional to $1/N_c$ while the width of Glueball  is proportional to $1/N_c^2$.
Another criterion is that   Glueball with non-zero spin ($J\neq 0$) is blind to quark flavor, while scalar Glueball with the quantum number
$0^{++}$ decays
to  $q\bar{q}$ is chiral suppressed~\cite{Chanowitz:2005du,Chao:2007sk} and thus its decay to $s\bar{s}$ is favored
than to $u\bar{u}$ or $d\bar{d}$.

The search of Glueball can be carried out in many experiments: $p\bar{p}$ collision, $\pi N$ scattering on
polarized/unpolarized targets, central hadronic production, $B$-meson decay,  and radiative decays of
vector states of heavy quarkonia,
i.e. $J/\psi$, $\psi(2S)$ and $\Upsilon(nS)$~\cite{Shen:2002nv,Bai:2003ww,Ablikim:2013hq,Dobbs:2015dwa}.
A great deal of data has been obtained and it is believed that a large possibility of Glueball component in
the scalar state  $f_0(1500)$ or  $f_0(1710)$~\cite{Cheng:2015iaa}. But a pure Glueball where only gluons contribute the
constituent  has not been observed or verified up to now.  The study of glue-rich  processes
shall be greatly helpful to hunt the signal of Glueball. The radiative decay of vector states of heavy quarkonia
through $V(1^{--})\to G+\gamma $
is one of the most important processes which shall provide a platform to systematically investigate the properties of
Glueball.

In the case of the mass squared of produced Glueball greatly less than that of the parent heavy quarkonium, i.e.
$m_G^2<<m_V^2$, a large momentum is transferred,
the final
Glueball and photon both run in the light-cone, and the light-cone factorization
can be well-employed. The soft and collinear physics is only contained in LCDA of Glueball and the NRQCD long-distance
matrix elements of the parent heavy quarkonium.
The light-cone operators populate the collinear subgroup of four-dimensional conformal symmetry,
and the LCDAs of Glueball which incorporate the gluon distribution with certain light-cone
momentum fraction can be defined  accordingly.
The scale dependence of the corresponding physical
observable is governed by evolution equation. The evolution equation of LCDAs can be understood as
the renormalization group equations for the light-cone operators. Some pioneer works on the evolution equation of distribution amplitudes
for exclusive reactions at large momentum transfer can be found in
Refs.~\cite{Lepage:1980fj,Efremov:1979qk,Duncan:1979hi,Ji:1996nm,Belitsky:1998gc,Ji:2002aa}. The study of Glueball from radiative Upsilon
decay based on soft-collinear effective theory (SCET) can be found in Ref.~\cite{Fleming:2004rk}.

In the paper, we establish the light-cone factorization for scalar Glueball production in the radiative $J/\psi$ decay,
which will be tested at next-to-leading order. The factorization formula is also valid in other vector heavy quarkonium decays,
e.g. $\psi(nS)$ and $\Upsilon(nS)$. Through the establishment of the factorization formulae,
it is conveniently to perform a systematical  phenomenology analysis and open
a new and clear widow to investigate the properties of Glueball.

We first give the definition of leading-order LCDAs for scalar Glueball. Since the Glueball and the flavor-singlet $q\bar{q}$ state have identical conformal spin, they will
mix each other by renormalization, which is ananogous to the quark-gluon splitting behavior in Proton. Thus we define a two dimensional light-cone distribution amplitude
\beq
\mbox{\boldmath$\Phi$}(u)\equiv \left(
\begin{array}[c]{c}
\phi_{q}(u) \\[0.2em]
\phi_{g}(u)
\end{array}
\right) \, ,\label{eq:definition-LCDA}
\eeq
where the twist-2 LCDAs of $\phi_{i}(u)$ can be written as follows in terms of quark and gluon
fields
\begin{eqnarray}
 \phi_q (u) &=& \int \frac{dz^-}{2\pi}\frac{ e^{i(2u-1)k^+z^-/2 } }{N_q }\langle G(k)|\overline{\Psi}_i(-z^-/2)L_{ij}(-z^-/2,z^-/2)\Psi_j(z^-/2)|0\rangle,
 \label{eq:definition-gauge-invariant-LCDA-q}
\end{eqnarray}
\begin{eqnarray}
 \phi_g (u) &=& \int \frac{dz^-}{2\pi}
\frac{ e^{i(2u-1)k^+z^-/2 } }{N_g u(1-u)}g_\perp^{\mu\nu}\langle G(k)|G^{a,+\mu}(-z^-/2)L_{ab}
(-z^-/2,z^-/2)G^{b,+\nu}(z^-/2)|0\rangle,~~
 \label{eq:definition-gauge-invariant-LCDA-g}
\end{eqnarray}
where the resummation of all order soft and collinear gluon radiation from quark or gluon field is summarized into
the related gauge link, which also ensure the gauge invariant of the defined matrix elements.  For a vector $p$, the light-cone component is given by $p^\mu=(p^+,p^-,p^1,p^2)$
with $p^+=(p^0+p^3)/\sqrt{2}$ and $p^-=(p^0-p^3)/\sqrt{2}$.  The $u$ defined above is the
momentum fraction in plus direction for one gluon in Glueball. The factor $N_i$ satisfies
$N_q=1$ and $N_g=k ^+$. We define two light-cone vectors $n$ and $\bar{n}$ with $n^\mu=(1,0,0,0)$ and
$\bar{n}^\mu=(0,1,0,0)$ in the light-cone frame. The tensor factor $g_\perp^{\mu\nu}$ can be written as $g_\perp^{\mu\nu}
=g^{\mu\nu}-n^\mu\bar{n}^\nu-n^\nu\bar{n}^\mu$.
For the gauge link, we have
\begin{eqnarray}
L(x,y) &=& P\,e^{ig\int_0^1 ds (x-y)_\mu A^\mu((x-y)s+y)},~~
\end{eqnarray}
where $A^\mu=A_a^\mu T^a$ in the fundamental representation for the links between quark and anti-quark;
while $A^\mu= if^{abc}A_b^\mu$ in the adjoint representation for the links between gluon and gluon.

\section{Factorization formulae}

We consider the radiative decay of $J/\psi$ in its rest-frame~\cite{Cakir:1994jf,He:2002hr,Melis:2004ni}
\beq
J/\psi(P)\to G(k)+\gamma(q)\,,
\eeq
where the related momenta are given in the brackets. The momentum of Glueball can
be written explicitly as $({\cal O}(k^+),{\cal O}(k^-),
{\cal O}(\Lambda_{QCD}),{\cal O}(\Lambda_{QCD}))$. Since the charm quark is heavy,
the produced Glueball has a momentum of order of $m_c$, and satisfies
\beq
\frac{k^-}{k^+}=\frac{m_G^2}{m_{J/\psi}^2}\sim 0.30,\qquad
\frac{\Lambda_{QCD}}{k^+}=\frac{\sqrt{2}\Lambda_{QCD}}{m_{J/\psi}}\sim 0.14,
\eeq
where we assume that $m_G\approx 1.7${GeV}, $\Lambda_{QCD}\approx0.3${GeV}. If the parent
heavy quarkonium becomes to $\Upsilon$, the corresponding ratios of $k^-/k^+$ and $\Lambda_{QCD}/k^+$
dramatically decrease to 0.03 and 0.04 respectively. Thus the light-cone factorization is more practical in
the $\Upsilon$ radiative decay.

We only consider the QCD corrections here, so the S-matrix element for the decay is
\beq
\langle \gamma(q) G(K)|S|J/\psi\rangle=-ieQ_c\varepsilon^{*\mu}(q)\int d^4x e^{iq\cdot x}
\langle G(k)|\bar{c}(x)\gamma_\mu c(x)|J/\psi(P)\rangle\,,
\eeq
where $Q_c$ is the electric charge of charm quark, $c(x)$ is the Dirac field for the charm quark, $\varepsilon^\mu$
is the polarization vector for the photon. At the leading-order twist defined in
Eq.~(\ref{eq:definition-gauge-invariant-LCDA-g}), at least two gluons bound to Glueball. The corresponding
contribution to the S-matrix element is
\bqa
\langle \gamma(q) G(K)|S|J/\psi\rangle&=&-i\frac{1}{2}e Q_c g_s^2 \varepsilon^{*\alpha}(q)\int d^4x d^4y d^4 ze^{iq\cdot x}\nnb
&&
\times\langle G(k)|T[\bar{c}(x)\gamma_\mu c(x)\bar{c}(y)\gamma\cdot G(y) c(y)\bar{c}(z)\gamma\cdot
G(z) c(z)]|J/\psi(P)\rangle\,,
\eqa
We can calculate the T-ordered operator product by Wick-contraction and we use the expansion of the heavy quark relative velocity $v$~\cite{Ma:2001tt}
\beq
\langle 0|\bar{c}_i(x) c_j (y)|J/\psi\rangle=-\frac{1}{6}\left(P^+\gamma_\mu P^-\right)_{ji}\langle 0|\chi^\dagger
\sigma^\mu \psi|J/\psi\rangle e^{-iP\cdot (x+y)}+{\cal O}(v^2)\,,
\eeq
where  $\psi$ and $\chi^\dagger$ are the NRQCD operators to annihilate the quark and anti-quark respectively. Since the heavy
quark relative velocity squared is around 0.3 for $J/\psi$ and 0.1 for $\Upsilon$\cite{Bodwin:1994jh}, we neglect
the contribution from higher orders of $v^2$ in this paper. By the simplification, the amplitude can be written as
\bqa
\langle \gamma(q) G(K)|S|J/\psi\rangle&=&i\frac{1}{24}e Q_c g_s^2 \varepsilon^{*\alpha}(q)(2\pi)^4\delta(P-k-q)\langle 0|\chi^\dagger
\sigma^\beta \psi|J/\psi\rangle\nnb
&&
\times\int\frac{d^4 q_1}{(2\pi)^4}\Gamma^{\mu\nu}(k,q_1)M_{\alpha\beta\mu\nu}(P,k,q_1)\,,\label{nfactorization}
\eqa
where
\beq
\Gamma^{\mu\nu}(k,q_1)=\int d^4x e^{-iq_1\cdot x-i(k-q_1)\cdot y}\langle G(k)|G^{a,\mu}(x)G^{a,\nu}(y)|0\rangle\,,
\eeq
the function $\Gamma^{\mu\nu}(k,q_1)$ incorporates the non-local interactions among two gluons and Glueball, while the function
$M_{\alpha\beta\mu\nu}(P,k,q_1)$ is a perturbative kernel, which can be calculated order by order. At twist-2 level the
function $\Gamma^{\mu\nu}(k,q_1)$ can be simplified into
\beq
\Gamma^{\mu\nu}(k,q_1)|_{twist-2}=(2\pi)^4\delta(q_1^-)\delta^2({q_{1}}_\perp)\frac{1}{u(u-1) }g^{\mu\nu}_\perp F_0(u)\,,
\eeq
\beq
F_0(u)=\frac{1}{2\pi(k^+)^2}\int dx^-e^{-i(1-2u)k^+x^-}\langle G(k)|G^{a}_{+\mu}(-x^-)G^{a}_{+\nu}(x^-)|0\rangle\,.
\eeq

The naive factorization in Eq.~(\ref{nfactorization}) is valid in tree-level. At NLO and beyond NLO, the factorization
should be corrected to including the fluctuation  between the gluonium component and the $q\bar{q}$ flavor-singlet component.

Leaving all possible Lorentz invariant construction, the amplitude of $J/\psi$ radiative decays to scalar Glueball can be factorized into
\bqa
iM&=&ieQ_c g_s^2 \langle 0|\chi^\dagger{\mbox{\boldmath $\sigma$}}\psi|J/\psi\rangle \nnb&&\times\int_0^1 du \int_0^1 dt ( \varepsilon_{J/\psi}
\cdot\varepsilon_\gamma
m_c^2 \mbox{\boldmath$H$}_0(u,v,\mu)+\varepsilon_{J/\psi}\cdot q\,\varepsilon_\gamma\cdot P \mbox{\boldmath$H$}_1(u,v,\mu))
\mbox{\boldmath$\Phi$}(t,\mu)\,,\nnb
\eqa
with
\bqa
\langle 0|\chi^\dagger{\mbox{\boldmath $\sigma$}}\psi|J/\psi\rangle&=&
\Gamma_{J/\psi}(v,\mu)\langle 0|\chi^\dagger{\mbox{\boldmath $\sigma$}}\psi|J/\psi\rangle^r\,,\\
\mbox{\boldmath$\Phi$}(t,\mu)&=&\mbox{\boldmath$\Gamma$}(u,t,\mu)
\mbox{\boldmath$\Phi$}^r(t,\mu)\,,
\eqa
Note that $\langle 0|\chi^\dagger{\mbox{\boldmath $\sigma$}}\psi|J/\psi\rangle^r$ here is the matrix element after
renormalization, which is isolated to the renormalization of LCDAs for Glueball. $\Gamma_{J/\psi}$ is the renormalization
factor of LDME for $J/\psi$. $\mbox{\boldmath$\Gamma$}$ is the renormalization factor with $2\times 2$ matrix elements,
which can be calculated
through the renormalization of LCDAs of Glueball. $\varepsilon_{J/\psi}$ and $\varepsilon_\gamma$ are the polarization
vectors of $J/\psi$ and the radiated photon, respectively.
$\mbox{\boldmath$H$}_i$ is the hard kernel
with two components. The factorization formula can also employed to $\Upsilon \to \gamma+G$ by
replacing $Q_c\to Q_b$, $m_c\to m_b$, $\varepsilon_{J/\psi}\to \varepsilon_\Upsilon$, and $m_{J/\psi}\to m_\Upsilon$.

\begin{figure}[th]
\begin{center}
\includegraphics[width=0.55\textwidth]{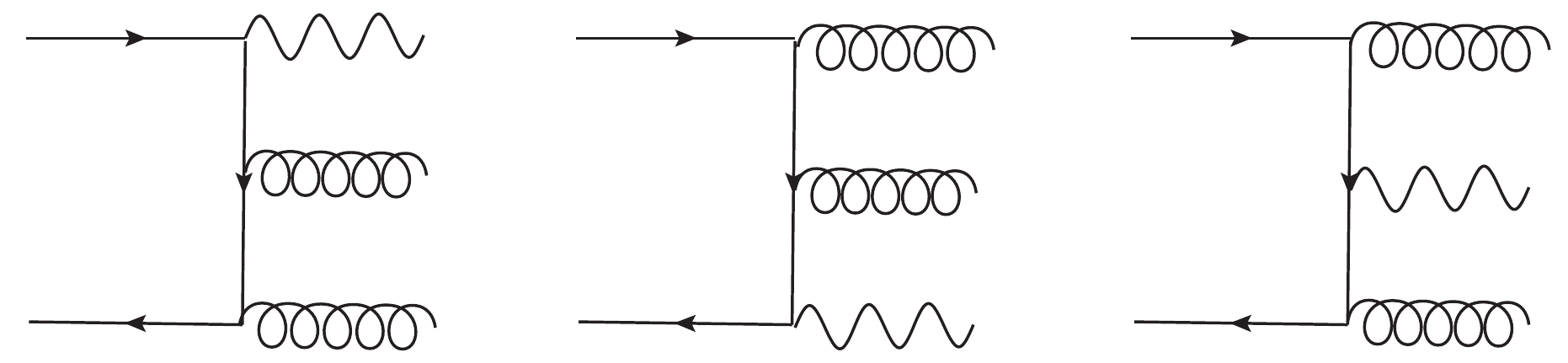}
\includegraphics[width=0.58\textwidth]{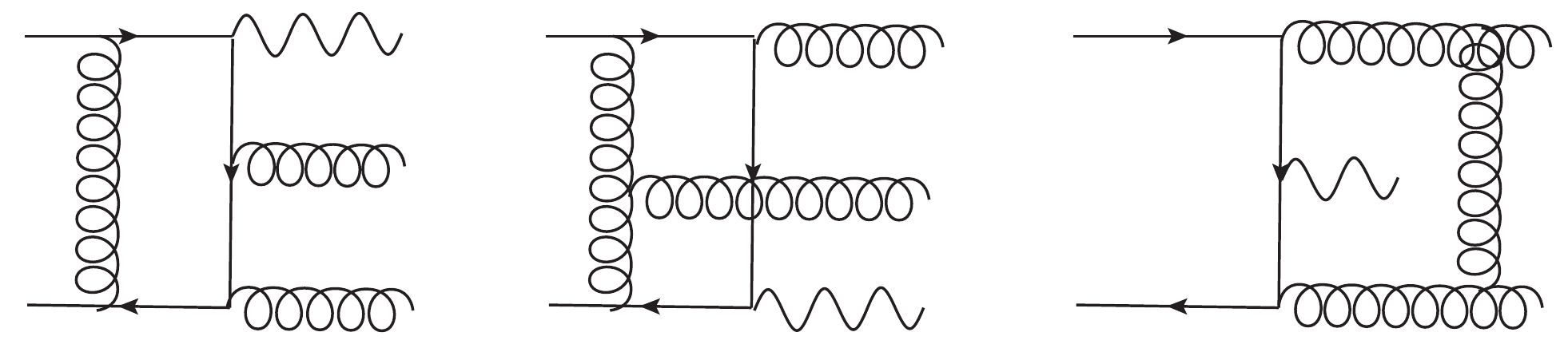}
\includegraphics[width=0.80\textwidth]{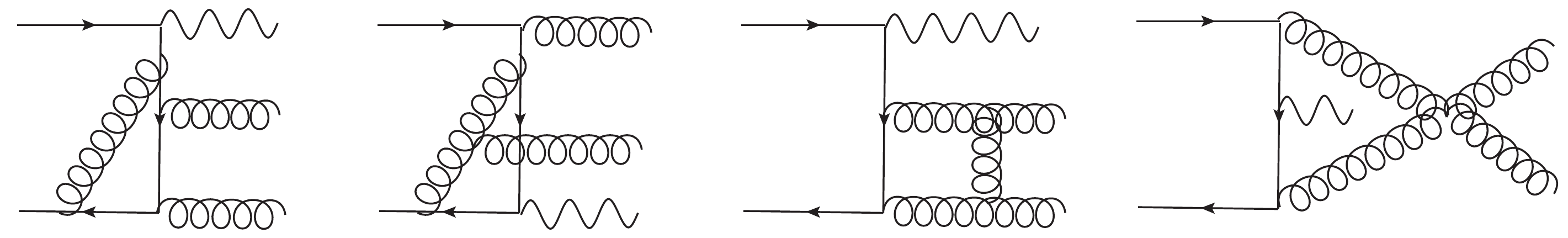}
\includegraphics[width=0.78\textwidth]{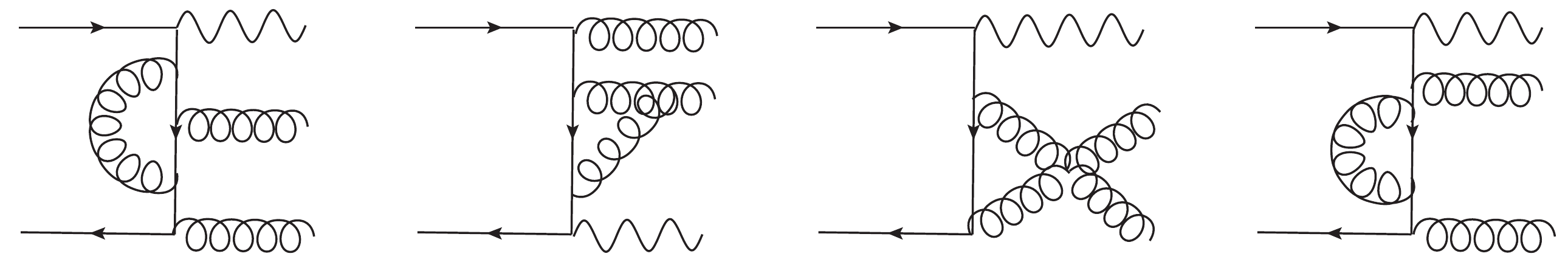}
\includegraphics[width=0.50\textwidth]{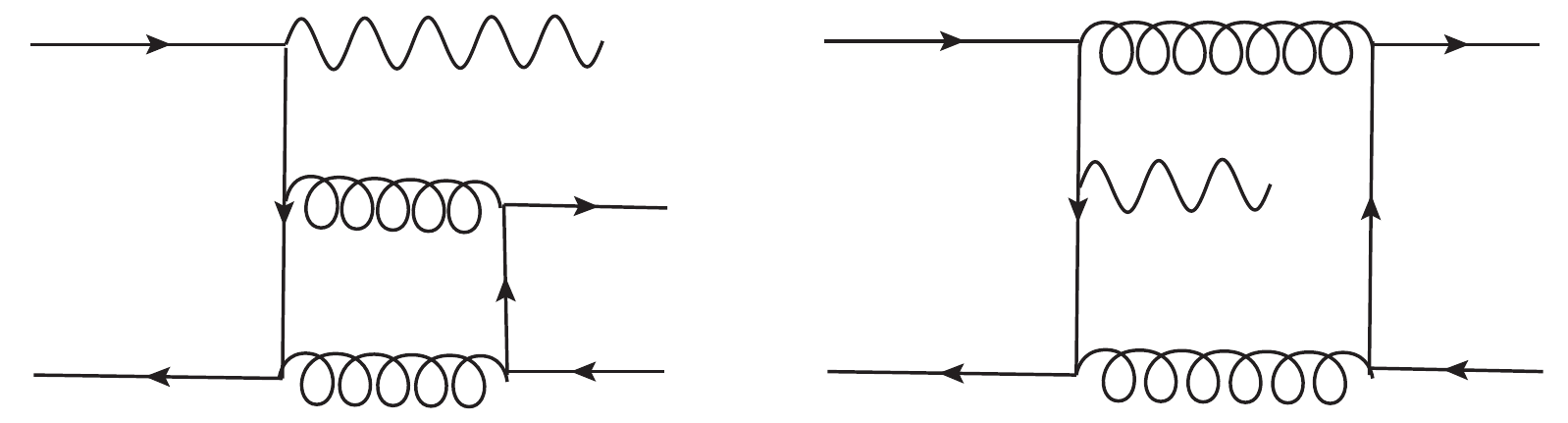}
\end{center}
    \vskip -0.7cm \caption{Typical Feynman diagrams for the hard kernel of  a heavy quarkonium radiative decays to Glueball.}\label{Fig-hardkernel}
\end{figure}

The typical Feynman diagrams at both tree and one-loop level contributing to the hard kernel for a heavy
quarkonium radiative decays to Glueball can be found in Fig.~\ref{Fig-hardkernel}.
Other 73 diagrams for $gg\gamma$ final states can be obtained by exchanging the outgoing gluons or inverting the quark line in Fig.~\ref{Fig-hardkernel}.
For tree level, other 3
symmetrical diagrams can be obtained by inverting the direction of the quark in the first line.
For one-loop level, there are another 5, 3, and 1 pentagon diagrams respectively compared with the typical diagrams in the second line.
There are another 11, 7, 3, and 0 box diagrams respectively in the third line. And there are another 17, 11, and 1 triangle
diagrams, another 11 self-energy diagrams  in the fourth line. While the diagrams in the fifth line denotes the contribution
to $q\bar{q}\gamma$ final states, and another 3 box diagrams and 1 pentagon diagrams are not shown.

After the calculation, we can easily get the LO hard kernel
\bqa
\mbox{\boldmath$H$}^{(0)}_0=\left(0,-\frac{\sqrt{2}}{N_c m_c^3\sqrt{m_{J/\psi}}u(1-u)}\right)\,,\quad
\mbox{\boldmath$H$}^{(0)}_1=\left(0,\frac{1}{N_c m_c^3\sqrt{2m_{J/\psi}}u\bar{u}}\right)\,.
\eqa
The NLO hard kernel can also be obtained after considering the
field renormalization and counter-term. It can be written as
\bqa
\mbox{\boldmath$H$}^{(1)}_0&=&\frac{1}{\epsilon}
\frac{\sqrt{2}\alpha_s}{ 2\pi N_c m_c^3\sqrt{m_{J/\psi}}u\bar{u}}\left(
\frac{n_f(2u-1)(u\ln u+\bar{u}\ln\bar{u})}{2u\bar{u}}-\epsilon\,n_f H^a_0,\right.\nnb&&\left.
\frac{C_A (2u^2-2u+1)(u\ln u+\bar{u}\ln\bar{u})}{u\bar{u}}+\frac{\beta_0}{2}
+\frac{C_F}{4v}(\pi^2\epsilon-i\pi(\frac{\mu}{2m_cv})^{2\epsilon})
-\epsilon\,H^b_0\right)\,,\nnb
\mbox{\boldmath$H$}^{(1)}_1&=&-\frac{1}{2}\mbox{\boldmath$H$}^{(1)}_0|_{H^a_0\to H^a_1,\,H^b_0\to H^b_1}\,,
\eqa
where the coefficients $H^a_0$, $H^b_0$ which contribute to the finite term is presented in the Appendix.

The renormalization factor $\Gamma_{J/\psi}$ can be obtained by the renormalization of the naive LDMEs
$\langle 0|\chi^\dagger{\mbox{\boldmath $\sigma$}}\psi|J/\psi\rangle$. At one-loop level,
it can be written as~\cite{Bodwin:1994jh}
\beq
\Gamma^{(0)}_{J/\psi}=1\,,\quad\quad\quad  \Gamma^{(1)}_{J/\psi}=\frac{\alpha_s C_F}{4 \pi v}\left(
\pi^2-i\pi(\frac{1}{\epsilon}+\ln\frac{\mu^2}{4m_c^2 v^2})\right)\,.
\eeq

The factor $\mbox{\boldmath$\Gamma$}$  can also be obtained by the renormalization of LCDAs for Glueball.
Calculating the LCDAs defined in Eq.~(\ref{eq:definition-LCDA}), we can easily get the tree-level result
\beq
\mbox{\boldmath$\Gamma$}^{(0)}(u,t,\mu)=\left(
\begin{array}{cc}
\delta(u-t)& 0\\[0.2cm]
0 & \delta(u-t)
\end{array}
\right)\,.
\eeq

\begin{figure}[th]
\begin{center}
\includegraphics[width=0.75\textwidth]{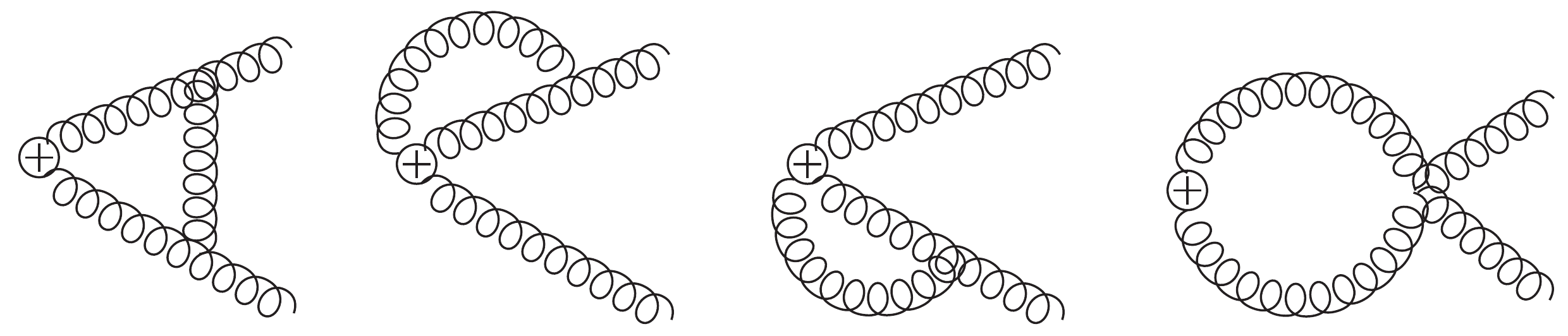}
\includegraphics[width=0.60\textwidth]{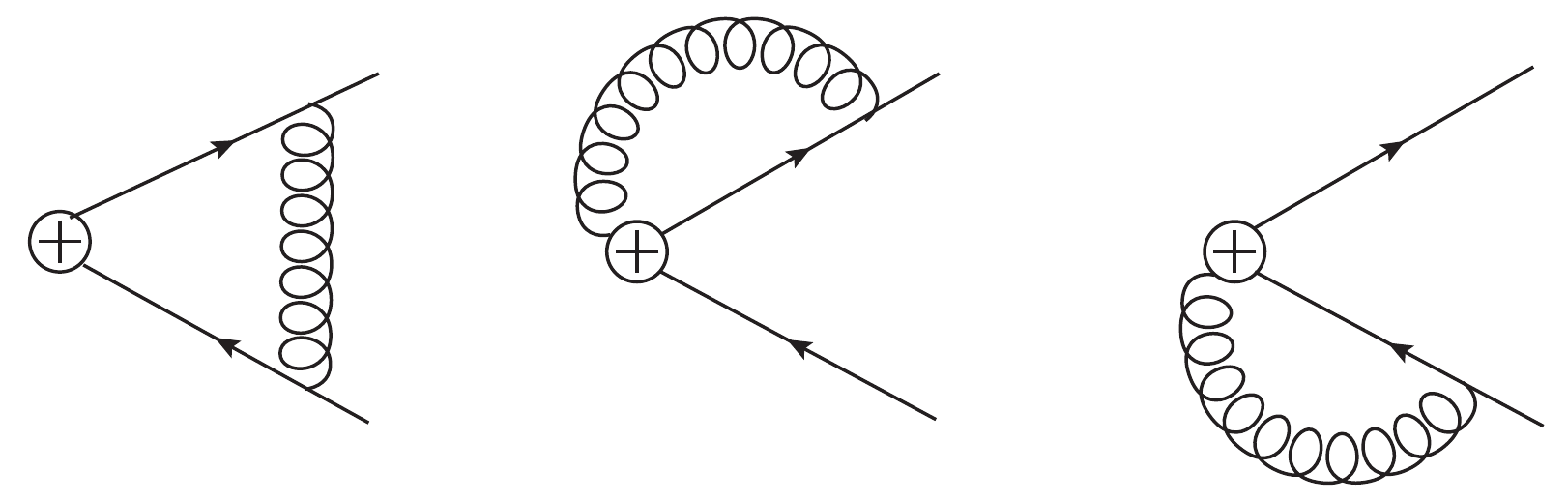}
\includegraphics[width=0.60\textwidth]{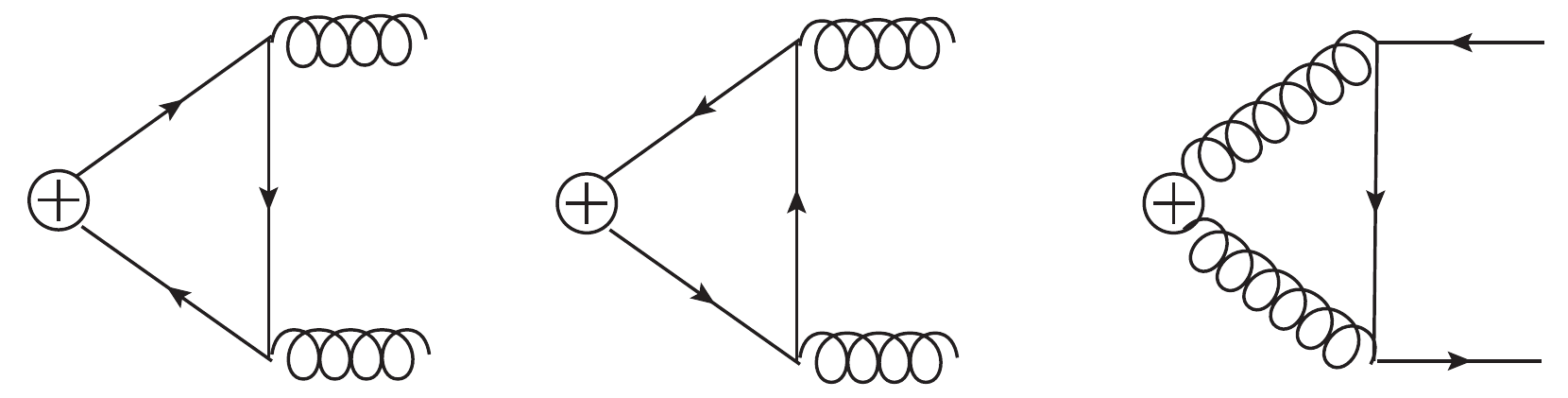}
\end{center}
    \vskip -0.7cm \caption{The renormalization of LCDAs for Glueball at one-loop level.}\label{Fig-soft}
\end{figure}

At one-loop level, the related Feynman diagrams can be found in Fig.~\ref{Fig-soft}, and we have
\beq
\mbox{\boldmath$\Gamma$}^{(1)}(u,t,\mu)=\frac{\alpha_s}{2\pi}(\frac{\mu}{\mu_0})^{2\epsilon}\frac{1}{\epsilon}\left(
\begin{array}{cc}
S^{(1)}_{qq}& S^{(1)}_{qg}\\[0.2cm]
S^{(1)}_{gq} & S^{(1)}_{gg}
\end{array}
\right)\,,
\eeq
with
\bqa
S^{(1)}_{qq}(u,t)&=&C_F\frac{u}{t}\left(1+\frac{1}{t-u}\right)_+ \theta(t-u)+\left(u\to\bar{u},t\to\bar{t}\right)\,,\nnb
S_{qg}^{(1)}(u,t)&=&2n_f T_F \frac{u}{t^2\bar{t}}\left(2u-t-1\right)\theta(t-u)-\left(u\to\bar{u},t\to\bar{t}\right)\,,\nnb
S^{(1)}_{gq}(u,t)&=&C_F\frac{u}{t}\left(2t-u\right)\theta(t-u)-\left(u\to\bar{u},t\to\bar{t}\right)\,,\nnb
S^{(1)}_{gg}(u,t)&=&C_A\frac{u^2}{t^2}\left(\frac{1}{(t-u)_+}+2
\left(\bar{u}+t(1+2\bar{u})\right)\right)\theta(t-u)\nnb
&&+\frac{\beta_0}{2}\delta(u-t)+\left(u\to\bar{u},t\to\bar{t}\right)\,,
\eqa
where $\beta_0=11N_c/3-2n_f/3$, the group factors $N_c=C_A=3$, $C_F=4/3$ and $T_F=1/2$ for $SU_c(3)$, and the plus function is defined as
\bqa
F(x,y)_+&=& F(x,y)-\delta(x-y)\int_0^1 dz F(z,y)\,.
\eqa

Note that the evolution kernel of LCDA is silimlar to that of the corresponding non-local operators, the latter ones have
been investigated in Refs.~\cite{Chase:1980hj,Belitsky:1998gc,Belitsky:1998vj}.

\section{Evolution equation for LCDAs of scalar Glueball}
The study of the universality of LCDAs for Glueball which describes
the long-distance interactions effects is a crucial and also interesting issue~\cite{Ma:2013yla,Wang:2013ywc}. One can find that all the divergences are cancelled out between the hard kernel and the renormalization factors for both LCDAs and NRQCD LDMEs.
We now turn to the scale evolution of LCDAs for Glueball.
The reason is that logarithms of the form $(\alpha_s \ln(m_{c,b}^2/\mu_0^2))^n$, where $\mu_0\sim 1{GeV}$ denotes
the scale at which nonperturbative physics of the LCDAs exists, are large and must be resummed to all orders.
By the renormalization equation, we can resum these large logarithms.  The scale dependent equation of LCDA for
Glueball reads
\beq
\mu^2\frac{\partial}{\partial\mu^2}\mbox{\boldmath$\Phi$}(u,\mu^2)=\mbox{\boldmath$V$}(u,t,\alpha_s(\mu^2))\otimes
\mbox{\boldmath$\Phi$}(t,\mu^2)\,,\label{req}
\eeq
where the evolution kernel $\mbox{\boldmath$V$}
$ is
\beq
\mbox{\boldmath$V$}=-\mbox{\boldmath$\Gamma$}^{-1}
\otimes\left(\mu^2\frac{\partial}{\partial\mu^2}\mbox{\boldmath$\Gamma$}\right)=\frac{\alpha_s(\mu^2)}{2\pi}
\left(
\begin{array}{cc}
S^{(1)}_{qq}& S^{(1)}_{qg}\\[0.2cm]
S^{(1)}_{gq} & S^{(1)}_{gg}
\end{array}
\right)+{\cal O}(\alpha_s^2)\,.
\eeq

The eigenfunctions of Eq.~(\ref{req}) after the  renormalization are the Gegenbauer polynomials and hence
the LCDAs of Glueball $\mbox{\boldmath$\Phi$}(u,\mu^2)$ possess the expansion~\cite{Fleming:2004rk,Kroll:2002nt}
\bqa
\Phi_q(u,\mu^2)&=&6u(1-u)f_q\sum_{n=1,3,...} a^q_n(\mu^2)C_n^{3/2}(2u-1)\,,
\nnb
\Phi_g(u,\mu^2)&=&30u^2(1-u)^2f_g\left(1+\sum_{n=3,5,...}  a^g_n(\mu^2)C_{n-1}^{5/2}(2u-1)\right)\,,
\eqa
where we omit the even $n$ series as a consequence of the symmetry of LCDA for scalar Glueball, i.e.
$\Phi_g(u,\mu^2)=\Phi_g(1-u,\mu^2)$ and $\Phi_q(u,\mu^2)=-\Phi_q(1-u,\mu^2)$. The Gegenbauer momentum $a_n$
also obey the renormalization group equation
\beq
\mu^2\frac{\partial}{\partial\mu^2} \left(
\begin{array}[c]{c}
a^q_{n}(\mu^2) \\[0.2em]
a^g_{n}(\mu^2)\end{array}
\right) =\frac{\alpha_s(\mu^2)}{2\pi}
\left(
\begin{array}{cc}
\gamma_n^{qq}& \gamma_n^{qg}\\[0.2cm]
\gamma_n^{gq} &\gamma_n^{gg}
\end{array}
\right)
 \left(
\begin{array}[c]{c}
a^q_{n}(\mu^2) \\[0.2em]
a^g_{n}(\mu^2)\end{array}
\right)\,,
\eeq
where the anomalous dimensions that govern the evolution of the LCDAs are
\bqa
\gamma_n^{qq}&=&C_F\left(3+\frac{2}{(n+1)(n+2)}-4\psi(n+2)-4\psi(1)\right)\,,
\nnb
\gamma_n^{qg}&=&\frac{24n_fT_F(n^2+3n+4)}{n(n+1)(n+2)(n+3)}\,,
\nnb
\gamma_n^{gq}&=&\frac{C_F(n^2+3n+4)}{3(n+1)(n+2)}\,,
\nnb
\gamma_n^{gg}&=&C_A\left(-4\psi(n+2)+4\psi(1)+\frac{\beta_0}{2C_A}-\frac{8(n^2+3n+3)}{n(n+1)(n+2)(n+3)}\right)\,,
\eqa
where $\psi(x)$ is the digamma function.

After solving the renormalizaiton group equation, the resummed Gegenbauer momenta read
\bqa
a_n^q(\mu^2)&=&\frac{1}{\delta\gamma_n^{gq}}\left( a_n^+(\mu_0^2)\left(\lambda_n^+ -\gamma_n^{gg}\right)\left[\frac{\alpha_s(\mu^2)}{\alpha_s(\mu_0^2)}
\right]^{2\lambda_n^+/\beta_0}
-a_n^-(\mu_0^2)\left(\lambda_n^- -\gamma_n^{gg}\right)\left[\frac{\alpha_s(\mu^2)}{\alpha_s(\mu_0^2)}\right]^{2\lambda_n^-/\beta_0}\right)\,,
\nnb
a_n^g(\mu^2)&=&\frac{1}{\delta}\left( a_n^+(\mu_0^2)
\left[\frac{\alpha_s(\mu^2)}{\alpha_s(\mu_0^2)}\right]^{2\lambda_n^+/\beta_0}
-a_n^-(\mu_0^2)\left[\frac{\alpha_s(\mu^2)}{\alpha_s(\mu_0^2)}
\right]^{2\lambda_n^-/\beta_0}\right)\,,\nnb
\eqa
\begin{figure}[th]
\begin{center}
\includegraphics[width=0.75\textwidth]{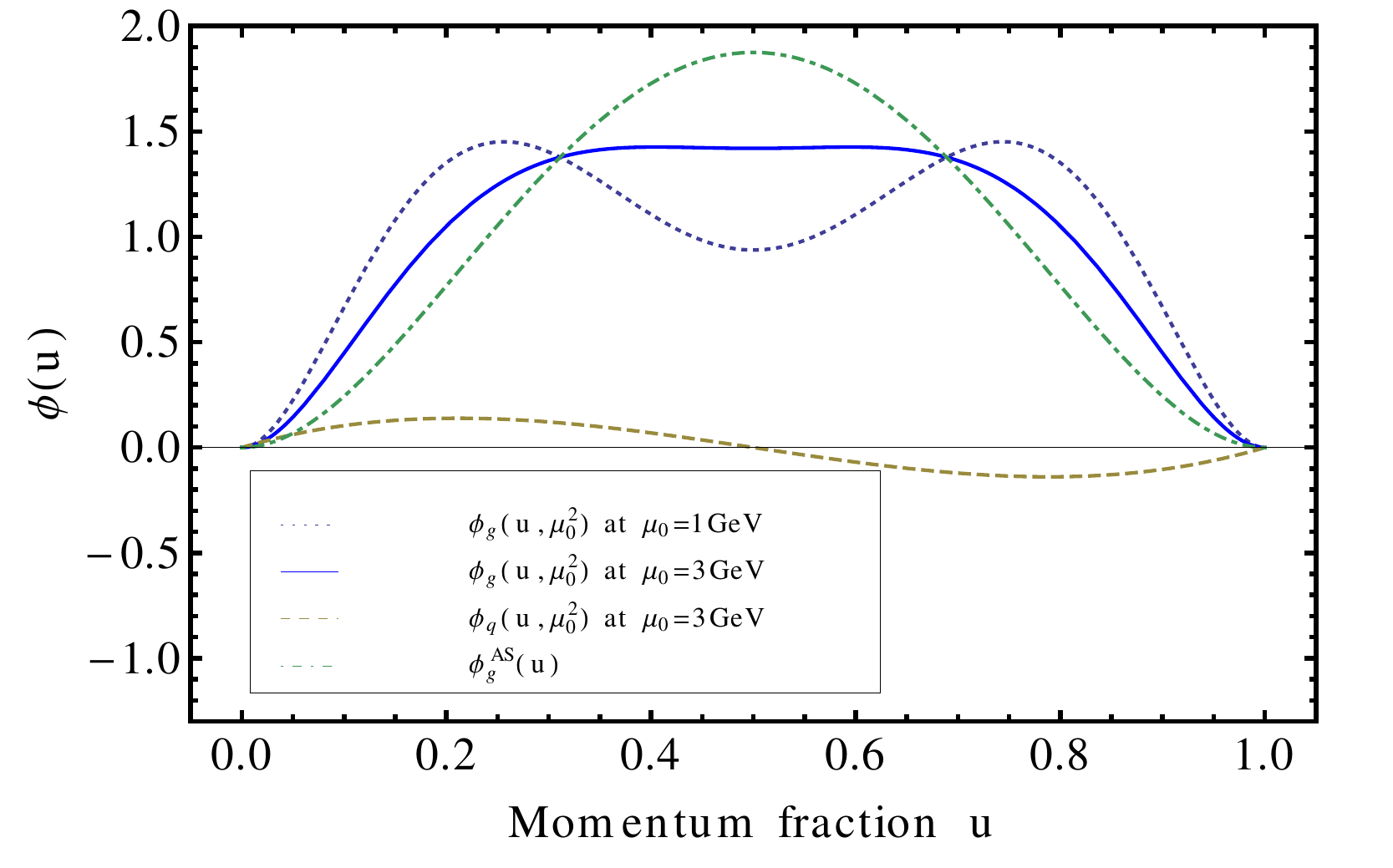}
\end{center}
    \vskip -0.7cm \caption{The light-cone distribution amplitude for scalar Glueball, where we only consider the first
    Gegenbauer momentum and resum the corresponding large logarithms in $a_3^g(\mu)$ and $a_1^q(\mu)$, using $\mu_0=1${GeV}
    and $a_3^g(\mu_0)=0.2$, $a_1^q(\mu_0)=0$ as input. The asymptotic form of
    $\phi_g (u)$ is renormalized to $30 u^2(1-u)^2$. }\label{Fig-lcda}
\end{figure}
where $\lambda_n^{\pm}$ are the eigenvalues
\beq
\lambda_n^{\pm}=\frac{1}{2}\left(\gamma_n^{gg}+\gamma_n^{qq}\pm\delta \right)\,,
\eeq
with $\delta=\sqrt{(\gamma_n^{gg}-\gamma_n^{qq})^2+4\gamma_n^{qg}\gamma_n^{gq}}$. And $a_n^{\pm}$ are
the eigenvectors, with
\beq
a_n^{\pm}(\mu^2)=a_n^{q}(\mu^2)\gamma_n^{gq}-a_n^{g}(\mu^2)(\lambda_n^{\mp}-\gamma_n^{gg})\,.
\eeq

We show  the LCDAs of scalar Glueball  in Fig.~\ref{Fig-lcda}, where we take $a_3^g(\mu_0)=0.2$ and $a_1^q(\mu_0)=0$ with $\mu_0=1${GeV} as input, and evolute
it into another scale.

\section{Phenomenological discussions\label{IV}}
In this section, we  will employ the above factorization formulae and analyze the phenomenological results
confronting the recent BESIII and CLEO data~\cite{Ablikim:2013hq,Dobbs:2015dwa}. The branching ratio of $J/\psi\to G+\gamma$ can be written as
\begin{eqnarray}
{\cal B}(J/\psi\to G+\gamma)&=&\frac{m_{J/\psi}^2-m_G^2}{16\pi\Gamma_{J/\psi}
m_{J/\psi}^3}|{\cal M}(J/\psi\to G+\gamma)|^2.
\end{eqnarray}
At first, we note that the decay width of $J/\psi\to G+\gamma$
has been studied by the CLQCD Collaboration within the framework of quenched Lattice QCD~\cite{Gui:2012gx}, which
gives
\begin{eqnarray}
{\cal B}(J/\psi\to G+\gamma)&=&(3.8\pm0.9)\times 10^{-3}.
\end{eqnarray}
In the following we will adopt the value of parameters from PDG2014:~\cite{Agashe:2014kda}
$m_{J/\psi}=3.0969$GeV, $\Gamma_{J/\psi}=92.9$keV, $m_{\psi(2S)}=3.686$GeV, $\Gamma_{\psi(2S)}=286$keV,
$m_{\Upsilon}=9.4603$GeV, $\Gamma_{\Upsilon}=54.02$keV, $m_{\Upsilon(2S)}=10.023$GeV, $\Gamma_{\Upsilon(2S)}=31.98$keV,
$m_{\Upsilon(3S)}=10.355$GeV,
$\Gamma_{\Upsilon(3S)}=20.32$keV. The values of LDMEs for heavy quarkonia are extracted
from their electric widths at NLO as Ref.~\cite{Zhu:2015jha}, which reads as
$\langle 0|\chi^\dagger{\mbox{\boldmath $\sigma$}}\psi|J/\psi\rangle=0.6408(GeV)^{3/2}$,
$\langle 0|\chi^\dagger{\mbox{\boldmath $\sigma$}}\psi|\psi(2S)\rangle=0.4975(GeV)^{3/2}$,
$\langle 0|\chi^\dagger{\mbox{\boldmath $\sigma$}}\psi|\Upsilon\rangle=1.710(GeV)^{3/2}$,
$\langle 0|\chi^\dagger{\mbox{\boldmath $\sigma$}}\psi|\Upsilon(2S)\rangle=1.2502(GeV)^{3/2}$
and $\langle 0|\chi^\dagger{\mbox{\boldmath $\sigma$}}\psi|\Upsilon(3S)\rangle=1.099(GeV)^{3/2}$. The heavy quark mass is adopted as $m_c=1.5${GeV}
and $m_b=4.8${GeV}~\cite{Qiao:2014pfa,Wang1}. From Fig.~\ref{Fig-lcda}, one can see that the $\Phi_q(u)$ is
small, so we will ignore its contribution in the following. We take the mass of scalar Glueball as $m_G=1.710$GeV from
Lattice QCD~\cite{Chen:2005mg}, then we can extract the decay constant of scalar Glueball, which reads as
\begin{eqnarray}
f_g=0.0386^{+0.0097}_{-0.0049}GeV,
\end{eqnarray}
where the uncertainty is from both the Lattice QCD and the running coupling constant. Note that the result is
a little smaller than the prediction $(0.10-0.13)${GeV} from QCD sum rule~\cite{He:2002hr}.
We will get a  more precise and reliable result for the decay constant of Glueball, if we have the Lattice
result for $\Upsilon(nS)\to G+\gamma$, where the NRQCD$\bigotimes$LCDA factorization becomes more solid.  We can also predict
the branching ratios of scalar Glueball from other vector heavy quarkonia. They are
\begin{eqnarray}
{\cal B}(\psi(2S)\to G+\gamma)&=&(5.9^{+3.4}_{-1.4})\times 10^{-4},\nonumber\\
{\cal B}(\Upsilon\to G+\gamma)&=&(1.3^{+0.7}_{-0.3})\times 10^{-4},\nonumber\\
{\cal B}(\Upsilon(2S)\to G+\gamma)&=&(1.0^{+0.6}_{-0.2})\times 10^{-4},\nonumber\\
{\cal B}(\Upsilon(3S)\to G+\gamma)&=&(1.2^{+0.7}_{-0.3})\times 10^{-4}.
\end{eqnarray}

Next we will consider the mixing among scalar Glueball and scalar $q\bar{q}$ states. There are many scalar mesons
with masses lower than 2GeV, which can be classified into two nonets: one nonet with mass below 1GeV includes $f_0(500)$,
$f_0(800)$, $K_0^*(800)$ and $a_0(980)$; the other nonet with mass above 1GeV includes $K_0^*(1430)$, $a_0(1450)$ and
two scalar mesons~\cite{Cheng:2015iaa,Cheng:2006hu}. One can see that not all three isosinglet scalars $f_0(1370)$, $f_0(1500)$ and $f_0(1710)$
can be accommodated in the $q\bar{q}$ nonet picture at the same time. One of them can have a large possibility of Glueball component.
Denoting $n\bar{n}=(u\bar{u}+d\bar{d})/\sqrt{2}$, we write the mixing formula~\cite{Close:2005vf,Cho:2015rsa}
\begin{eqnarray}
|f_0^i\rangle&=&\alpha_i|n\bar{n}\rangle+\beta_i|s\bar{s}\rangle+\rho_i|G\rangle,
\end{eqnarray}
where $f_0^i$ can be one of $f_0(1370)$, $f_0(1500)$ and $f_0(1710)$.

According to the lattice calculations~\cite{Lee:1999kv}, Lee and Weingarten found that $f_0(1710)$
is composed mainly of scalar Glueball. It is reasonable when we take the current experimental data into
account. From PDG2014, we find that ${\cal B}(\psi(J/\psi)\to f_0(1710) \gamma)$ is a large value,
which will be $1.56\times 10^{-3}$
when considering ${\cal B}(\psi(J/\psi)\to f_0(1710) \gamma\to \pi\bar{\pi}\gamma)=(4.0\pm1.0)
\times 10^{-4}$, ${\cal B}(\psi(J/\psi)\to f_0(1710) \gamma\to K\bar{K}\gamma)=(8.5^{+1.2}_{-0.9})
\times 10^{-4}$ and ${\cal B}(\psi(J/\psi)\to f_0(1710) \gamma\to \omega\bar{\omega}\gamma)=(3.1\pm1.0)
\times 10^{-4}$. On the other hand,  the fraction ${\cal B}(\psi(J/\psi)\to f_0(1710) \gamma)/
{\cal B}(\psi(J/\psi)\to f_0(1500) \gamma) $  is around order of 10,  and 
$f_0(1370)$ is still not observed in the $J\psi$ radiative 
decays.
The mixing matrix can be expressed as~\cite{Lee:1999kv}
\begin{eqnarray}
\left(
\begin{array}{c}
f_0(1370)\\[0.2cm]
f_0(1500)\\[0.2cm]
f_0(1710)
\end{array}
\right)&=&\left(
\begin{array}{ccc}
0.819(89)& 0.290(91)&-0.495(118)\\[0.2cm]
-0.399(113)& 0.908(37)&-0.128(52)\\[0.2cm]
0.413(87)& 0.302(52)&0.859(54)
\end{array}
\right)\,\left(
\begin{array}{c}
|n\bar{n}\rangle\\[0.2cm]
|s\bar{s}\rangle\\[0.2cm]
|G\rangle
\end{array}
\right). \label{mixing angle}
\end{eqnarray}
One can easily see that $f_0(1370)$ has a large possibility of the $n\bar{n}$ component, while $f_0(1500)$ is dominated
by the $s\bar{s}$ component.

We assume that Glueball component dominates the contribution in $V\to f_0^i +\gamma$ since  the processes from $q\bar{q}$ components are suppressed by the strong coupling squared $\alpha_s^2$.
Here we simply generalize it to all the three scalar mesons. The related branching ratios are given in Tab.~\ref{tab:data}, where one
can see that our results
 are comparable with data from PDG2014 except the predictions for $f_0(1370)$.  Employing the mixing matrix elements of Eq.~(\ref{mixing angle})
 based on Lattice QCD, we predict a large branching ratio for $V(1^{--})\to f_0(1370)+\gamma$ while there is
 no signal at experiment. We conclude that the first line in the matrix of Eq.~(\ref{mixing angle}) may be
 not precise enough and it  need to be checked by the following experiment.

\begin{table}[thb]
\caption{\label{tab:data} The branching ratios ($10^{-4}$) of $V(1^{--})\to f_0^i +\gamma$ and $V(1^{--})\to G(0^{++}) +\gamma$, where V denotes one of heavy quarkonium
$J/\psi$, $\psi(2S)$ and $\Upsilon(nS)$, $f_0^i$ denotes one of $f_0(1370)$, $f_0(1500)$ and $f_0(1710)$.}
\begin{center}
\begin{tabular}{cccccc}
\hline\hline
Branching ratio ($10^{-4}$)& This work & LQCD~\cite{Gui:2012gx}& He et al.~\cite{He:2002hr} &Cheng et al.~\cite{Cheng:2006hu}&   PDG2014~\cite{Agashe:2014kda}\\
\hline
${\cal B}(J/\psi\to f_0(1370)+\gamma)$~\footnote{Here we adopt the mixing matrix as Eq.~(\ref{mixing angle}) from Lattice QCD, however the mixing
matrix elements for $f_0(1370)$ we think still need to be tested further.} & $9.3\pm2.2$ & --& --& --& -- \\
${\cal B}(J/\psi\to f_0(1500)+\gamma)$ & $0.62\pm0.15$ & --& --& 2.9& $1.01\pm0.32$ \\
${\cal B}(J/\psi\to f_0(1710)+\gamma)$ & $28.0\pm 6.6$ & --& --& 14.5&  $>15.6$\\
${\cal B}(J/\psi\to G(0^{++})+\gamma)$ & $38\pm9$~\footnote{We use the Lattice QCD result to extract
the decay constant of Glueball.}& $38\pm9$ & --& --&  --\\
${\cal B}(\psi(2S)\to f_0(1370)+\gamma)$  & $1.45^{+0.83}_{-0.34}$& -- & --& --& --  \\
${\cal B}(\psi(2S)\to f_0(1500)+\gamma)$  & $0.97^{+0.56}_{-0.23}$ & --& --& --&  -- \\
${\cal B}(\psi(2S)\to f_0(1710)+\gamma)$  & $4.4^{+2.5}_{-1.0}$ & --& --& --&  $>0.9$ \\
${\cal B}(\psi(2S)\to G(0^{++})+\gamma)$  & $5.9^{+3.4}_{-1.4}$ & --& --& --&  -- \\
${\cal B}(\Upsilon\to f_0(1370)+\gamma)$  & $0.32^{+0.18}_{-0.08}$ & --& 4.8& --& --  \\
${\cal B}(\Upsilon\to f_0(1500)+\gamma)$  & $0.021^{+0.012}_{-0.005}$& -- & 4.2& --&  $<0.15$ \\
${\cal B}(\Upsilon\to f_0(1710)+\gamma)$  & $0.96^{+0.55}_{-0.23}$ & --& 1.5& --&  $<2.6$ \\
${\cal B}(\Upsilon\to G(0^{++})+\gamma)$  & $1.3^{+0.7}_{-0.3}$& --& -- & --&  $<2.6$ \\
${\cal B}(\Upsilon(2S)\to f_0(1370)+\gamma)$  & $0.26^{+0.14}_{-0.06}$ & --& --& --& --  \\
${\cal B}(\Upsilon(2S)\to f_0(1500)+\gamma)$  & $0.016^{+0.009}_{-0.004}$& -- & --&  --& -- \\
${\cal B}(\Upsilon(2S)\to f_0(1710)+\gamma)$  & $0.77^{+0.44}_{-0.18}$& -- & --& --&  $<5.9$ \\
${\cal B}(\Upsilon(2S)\to G(0^{++})+\gamma)$  & $1.0^{+0.6}_{-0.2}$ & --& --& --&  $<5.9$ \\
${\cal B}(\Upsilon(3S)\to f_0(1370)+\gamma)$  & $0.30^{+0.17}_{-0.07}$ & --& --& --& --  \\
${\cal B}(\Upsilon(3S)\to f_0(1500)+\gamma)$  & $0.019^{+0.010}_{-0.005}$& -- & --&  --& -- \\
${\cal B}(\Upsilon(3S)\to f_0(1710)+\gamma)$  & $0.90^{+0.51}_{-0.21}$ & --& --&  --& -- \\
${\cal B}(\Upsilon(3S)\to G(0^{++})+\gamma)$  & $1.2^{+0.7}_{-0.3}$ & --& --&  --& -- \\
\hline\hline
\end{tabular}
\end{center}
\end{table}

\section{Conclusion}

In this paper, we have established the factorization formulae for heavy vector quarkonium radiative decays into scalar Glueball, by studying one-loop corrections
to the hard kernel, LDMEs of heavy quarkonium, and LCDA of scalar Glueball.  The NRQCD$\bigotimes$LCDA factorization formulae shall be valid to all orders of the strong coupling
constant $\alpha_s$ in the leading-order twist and heavy quark velocity, after considering a two-dimensional LCDA of Glueball which is defined by non-local light-cone
gauge-invariant operators matrix elements. The universality of LCDA for Glueball ensures us to extract its
decay constant. Matching to the CLQCD results, we have extracted the decay constant
for scalar Glueball, i.e. $f_g=0.0386^{+0.0097}_{-0.0049}${GeV}. We also predict the branching ratios
of scalar Glueball from other heavy vector quarkonia such as $\psi(2S)$ and $\Upsilon(nS)$, which
 can be checked in the upcoming experiment. The factorization formulae can also
be applied to pseudoscalar and tensor Glueball production from  heavy vector quarkonia. A systematic study
on Glueball production and decay with different quantum numbers shall be investigated in order to hunting and identifying Glueball with a large confidence. We will address these issues in following studies.

\section*{Acknowledgments}
I greatly thank Profs. Xiangdong Ji, Cong-Feng Qiao, Feng Yuan, Jian-Ping Ma, and Chengping Shen
for fruitful discussions. I also specially thank Wei Wang for pointing out the phenomenological analyses.
This work was supported in part by a key laboratory grant from the Office of Science and Technology,
Shanghai Municipal Government (No. 11DZ2260700),
by Shanghai Natural  Science Foundation  under Grant No.15ZR1423100,
and by the Open Project Program of State Key Laboratory of Theoretical Physics,
Institute of Theoretical Physics, Chinese
Academy of Sciences, China (No.Y5KF111CJ1).

\section*{Appendix}
\begin{appendix}
In the appendix, we give the explicit results of the short-distance coefficients for hard kernels.
\bqa
H^a_0&=&[\frac{(2 u-1) N_c\log (u)}{6 (u-1)}\log\frac{\mu^2}{m_c^2}-\frac{(2 u-1) N_c (\text{Li}_2(1-2 u)+\log (u) (\log (u)-2+2 \log (2)))}{6 (u-1)}\nnb&&+
\frac{(2 u-1) \left(54-\left(\pi ^2-18-2f_1^a\right) N_c\right)}{144 (u-1) u}+\frac{1}{8(u-1)u}((2u-1)(3B_1-2B_2+3C_1
\nnb&&-(8u^2-8u+3)C_2)
+2(8u^2-9u+3)B_3)]-u\to(1-u)\,,\nnb
H^a_1&=&H^a_0\,,
\eqa
\bqa
H^b_0&=&[\frac{\left(2 u^2-2 u+1\right) N_c \log (u)}{u-1}\log\frac{\mu^2}{m_c^2}
-\frac{\left(2 u^2-2 u+1\right) N_c }{u-1}(\text{Li}_2(1-2 u)+\log (u) (\log (u)-2
\nnb&&+2 \log (2))-\frac{f_1^a}{12u})+\frac{2f^b_1+b^b_1B_1}{96 (u-1)^2
   u^2}+\frac{b^b_2B_2}{96 (u-1) u}+\frac{b^b_3B_3}{48 (u-1)
   u^2 (2 u-1)}+\frac{b^b_4B_4}{12 (u-1)^2}\nnb&&
   +\frac{c^b_1 C_1}{192 (1-2 u)^4 (u-1)^2 u^2}
   +\frac{c^b_2 C_2}{48 (u-1)^2 u^2}+\frac{c^b_3 C_3}{32 (1-2 u)^4
   (u-1) u^2}+\frac{c^b_4 C_4}{24 (u-1)^2 u}\nnb&&+\frac{c^b_5 C_5}{96 (1-2 u)^4 u^2}
   +\frac{c^b_6 C_6}{48 (1-2 u)^4 (u-1)^2}
   -\frac{\left(u^4-2 u^3+u^2-3\right) C_F}{2 (u-1)^2 u^2}C_7\nnb&&
   -\frac{\left(2 u^3-u^2-9 u+1\right) C_F}{4 (u-1) u^2}C_8]+u\to(1-u)\,,
\eqa
where
\bqa
f_1^a&=&12(\text{Li}_2(1-2 u)+\text{Li}_2(2 u)+\log(1-2u) (\log(u)+\log(2))-\pi^2\,,\nnb
f^b_1&=&((438-4 \pi ^2) u^4+(8 \pi ^2-876) u^3+(585-6 \pi ^2) u^2
+(2 \pi ^2-147) u-6) N_c\nnb&&-12
   (52 u^4-104 u^3+46 u^2+6 u-3) C_F-90 u^4+180 u^3-433 u^2+343 u-36\,,\nnb
b^b_1&=&3 (180 u^4-360 u^3+237 u^2-57 u-2) N_c+12 (2 u^4-4 u^3+8 u^2-6 u+3)
C_F\nnb&&-1172 u^4+2344 u^3-1587 u^2+415 u-36\,,\nnb
b^b_2&=&-15 (1-2 u)^2 N_c+12 C_F-124 u^2+124 u-45\,,\nnb
b^b_3&=&3 (1-2 u)^2 \left(48 u^3-92 u^2+63 u-10\right) N_c+12 (6 u^4-37 u^3
+47 u^2-21 u+3) C_F\nnb&&+1312 u^4-2588 u^3+1740 u^2-455 u+36\,,\nnb
b^b_4&=&\left(6 u^2-6 u+9\right) N_c-3 \left(4 u^2+3 u-1\right) C_F-2 u^2+3 u-1\,,
\eqa
\bqa
c^b_1&=&3 \left(1792 u^8-7168 u^7+12688 u^6-12976 u^5+8368 u^4-3472 u^3
+895 u^2-127 u+8\right) N_c\nnb&&+4 (u-1) u (1-2 u)^4 \left(51 C_F-40 u^2+40
   u-1\right)\,,\nnb
c^b_2&=&-3 \left(16 u^4-32 u^3+27 u^2-11 u+2\right) (1-2 u)^2 N_c
+6 (1-2 u)^2 C_F\nnb&&+u \left(100 u^3-200 u^2+117 u-17\right)\,,\nnb
c^b_3&=&u \left(-128 u^7+184 u^6+932 u^5-2372 u^4+2224 u^3-1039 u^2+246 u-23\right) N_c\nnb&&
-4 (1-2 u)^4 \left(3 u^3-4 u^2-5 u-1\right) C_F\,,\nnb
c^b_4&=&3 \left(10 u^3-24 u^2+8 u+7\right) C_F+u \left(-3 \left(u^2-3 u+4\right) N_c-32 u^3+37 u^2+4 u-9\right)\,,\nnb
c^b_5&=&3 \left(128 u^7-648 u^6+1356 u^5-1256 u^4+472 u^3-3 u^2-40 u+8\right) N_c\nnb&&
+12 \left(3 u^2-20 u+24\right) (1-2 u)^4 C_F-4 u (9u+2) (1-2
   u)^4\,,\nnb
c^b_6&=&(u-1) \left(3 \left(128 u^6-56 u^5-220 u^4+232 u^3-72 u^2+7 u+1\right) N_c
+16 (u-1) (1-2 u)^4\right)\nnb&&-12 (1-2 u)^4 \left(3 u^2+7 u+16\right)
   C_F\,,\nnb
\eqa
and
\bqa
H^b_1&=&H^b_0+[\frac{\tilde{c}^b_1 C_1}{96 (1-2 u)^4 (u-1) u}
   +\frac{\tilde{c}^b_3 C_3}{96 (1-2 u)^4 (u-1) u}+\frac{\tilde{c}^b_4 C_4}{24 (u-1) u}
   +\frac{\tilde{c}^b_5 C_5}{48 (1-2 u)^4 u^2}\nnb&&
   +\frac{\tilde{c}^b_6 C_6}{24 (1-2 u)^4 (u-1)^2}
   -\frac{\left(2 u^4-4 u^3+u^2+u-3\right) C_F}{(u-1)^2 u^2}C_7\nnb&&
  +\frac{\left(4 u^2-3 u+5\right) C_F}{2 (u-1) u}C_8+u\to(1-u)]\,,
\eqa
\bqa
\tilde{c}^b_1&=&-3 \left(352 u^6-1056 u^5+1752 u^4-1744 u^3+992 u^2-296 u+41\right) N_c
\nnb&&+24 \left(4 u^2-4 u+3\right) (1-2 u)^4 C_F+4 \left(16 u^2-16
   u+45\right) (1-2 u)^4\,,\nnb
\tilde{c}^b_3&=&12 (1-2 u)^4 \left(2 u^2+3 u+1\right) C_F-(u-1) (3 (176 u^6-496 u^5
+796 u^4\nnb&&-792 u^3+456 u^2-129 u+15) N_c-4 (1-2 u)^4 \left(8
   u^2-5 u-3\right))\,,\nnb
\tilde{c}^b_4&=&u \left(-3 (4 u-5) N_c+32 u^2-28 u-45\right)
-12 \left(2 u^3-5 u^2+3\right) C_F\,,\nnb
\tilde{c}^b_5&=&u (3 \left(176 u^6-560 u^5+988 u^4-1016 u^3+584 u^2-183 u+28\right) N_c\nnb&&
-2 (1-2 u)^4 \left(16 u^2+10 u-55\right))+12 \left(4 u^2-15
   u+12\right) (1-2 u)^4 C_F\,,\nnb
\tilde{c}^b_6&=&(u-1) (3 \left(176 u^6-496 u^5+796 u^4-792 u^3+456 u^2-129 u+15\right) N_c\nnb&&
-4 (1-2 u)^4 \left(8 u^2-5 u-3\right))-72 (1-2 u)^4
   \left(u^2+1\right) C_F\,.
   \eqa
The coefficients $b_i$, $c_i$ and $\tilde{c}_i$  are related to the
scalar Passarino-Veltman integrals defined
in Ref.~\cite{Passarino:1978jh,Hahn:1998yk}, and
here we have  the relation $C_i=m_c^2\,C^0_i$ :
\begin{eqnarray}
B_1&=&B_0\left(0,m_c^2,m_c^2\right),\nonumber\\
B_2&=&B_0\left(-m_c^2,0,m_c^2\right),\nonumber\\
B_3&=&B_0\left((1-2 u) m_c^2,0,m_c^2\right),\nonumber\\
B_4&=& B_0\left(4 u
   m_c^2,m_c^2,m_c^2\right),\nonumber\\
C^0_1&=&\text{C}_0\left(-m_c^2,m_c^2,0,m_c^2,0,m_c^2\right),\nonumber\\
C^0_2&=&\text{C}_0\left(0,(1-2 u) m_c^2,(2 u-1)
   m_c^2,m_c^2,m_c^2,0\right),\nonumber\\
C^0_3&=&\text{C}_0\left(m_c^2,4 u m_c^2,(2 u-1) m_c^2,0,m_c^2,m_c^2\right),\nonumber\\
C^0_4&=&\text{C}_0\left(m_c^2,0,(1-2 u)
   m_c^2,0,m_c^2,m_c^2\right),\nonumber\\
C^0_5&=&\text{C}_0\left(-m_c^2,0,(1-2 u) m_c^2,0,m_c^2,m_c^2\right),\nonumber\\
C^0_6&=&\text{C}_0\left(0,0,4 u
   m_c^2,m_c^2,m_c^2,m_c^2\right),\nonumber\\
C^0_7&=&\text{C}_0\left(4 m_c^2,0,0,m_c^2,m_c^2,m_c^2\right),\nonumber\\
C^0_8&=&\text{C}_0\left(4 m_c^2,0,4 u
   m_c^2,m_c^2,m_c^2,m_c^2\right).\nonumber\\
\end{eqnarray}
\end{appendix}


\begin{thebibliography}{99}
\bibitem{Albanese:1987ds}
  M.~Albanese {\it et al.}  [APE Collaboration],
  Phys.\ Lett.\ B {\bf 192}, 163 (1987).



\bibitem{Morningstar:1999rf}
  C.~J.~Morningstar and M.~J.~Peardon,
  Phys.\ Rev.\ D {\bf 60}, 034509 (1999)
  [hep-lat/9901004].



\bibitem{Chen:2005mg}
  Y.~Chen, A.~Alexandru, S.~J.~Dong, T.~Draper, I.~Horvath, F.~X.~Lee, K.~F.~Liu and N.~Mathur {\it et al.},
  Phys.\ Rev.\ D {\bf 73}, 014516 (2006)
  [hep-lat/0510074].

\bibitem{Gregory:2012hu}
  E.~Gregory, A.~Irving, B.~Lucini, C.~McNeile, A.~Rago, C.~Richards and E.~Rinaldi,
  JHEP {\bf 1210}, 170 (2012)
  [arXiv:1208.1858 [hep-lat]].


\bibitem{Bagan:1990sy}
  E.~Bagan and T.~G.~Steele,
  Phys.\ Lett.\ B {\bf 243}, 413 (1990).
\bibitem{Liu:1993it}
  J.~P.~Liu and D.~H.~Liu, J.\ Phys.\ G {\bf 19}, 373 (1993); W.~Shuiguo, Z.~Zhenyu and L.~Jueping,
  Phys.\ Rev.\ D {\bf 82}, 016003 (2010)
  [arXiv:1007.2465 [hep-ph]].
\bibitem{Huang:1998wj}
  T.~Huang, H.~Y.~Jin and A.~L.~Zhang,
  Phys.\ Rev.\ D {\bf 59}, 034026 (1999)[hep-ph/9807391];

\bibitem{Qiao:2014vva}
  C.~F.~Qiao and L.~Tang,
  Phys.\ Rev.\ Lett.\  {\bf 113}, no. 22, 221601 (2014)
  [arXiv:1408.3995 [hep-ph]]; X.~H.~Yuan and L.~Tang,
  Commun.\ Theor.\ Phys.\  {\bf 54}, 495 (2010)[arXiv:0911.0806 [hep-ph]]; G.~Hao, C.~F.~Qiao and A.~L.~Zhang,
  Phys.\ Lett.\ B {\bf 642}, 53 (2006)
  [hep-ph/0512214].
\bibitem{Csaki:1998qr}
  C.~Csaki, H.~Ooguri, Y.~Oz and J.~Terning,
  JHEP {\bf 9901}, 017 (1999)
  [hep-th/9806021].

\bibitem{Brunner:2015oqa}
  F.~Br¨¹nner, D.~Parganlija and A.~Rebhan,
  Phys.\ Rev.\ D {\bf 91}, no. 10, 106002 (2015)[arXiv:1501.07906 [hep-ph]]; F.~Br¨¹nner and A.~Rebhan,
  arXiv:1504.05815 [hep-ph]; D.~Parganlija,
  Acta Phys.\ Polon.\ Supp.\  {\bf 8}, no. 1, 219 (2015)[arXiv:1503.00550 [hep-ph]].
\bibitem{Sonnenschein:2015zaa}
  J.~Sonnenschein and D.~Weissman,
  arXiv:1507.01604 [hep-ph].

\bibitem{Carlson:1982er}
  C.~E.~Carlson, T.~H.~Hansson and C.~Peterson,
  Phys.\ Rev.\ D {\bf 27}, 1556 (1983).



\bibitem{Cheng:2015iaa}
  H.~Y.~Cheng, C.~K.~Chua and K.~F.~Liu,
  arXiv:1503.06827 [hep-ph].

\bibitem{He:2015owa}
  X.~G.~He and T.~C.~Yuan,
  Eur.\ Phys.\ J.\ C {\bf 75}, no. 3, 136 (2015)
  [arXiv:1503.03577 [hep-ph]].
\bibitem{Wang:2009rc}
  W.~Wang, Y.~L.~Shen and C.~D.~Lu,
  J.\ Phys.\ G {\bf 37}, 085006 (2010)
  [arXiv:0908.2216 [hep-ph]].


\bibitem{Chanowitz:2005du}
  M.~Chanowitz,
  Phys.\ Rev.\ Lett.\  {\bf 95}, 172001 (2005)
  [hep-ph/0506125].
\bibitem{Chao:2007sk}
  K.~T.~Chao, X.~G.~He and J.~P.~Ma,
  Phys.\ Rev.\ Lett.\  {\bf 98}, 149103 (2007)
  [arXiv:0704.1061 [hep-ph]].


\bibitem{Shen:2002nv}
  X.~Y.~Shen [BES Collaboration],
  eConf C {\bf 020620}, THAT07 (2002)
  [hep-ex/0209031].

\bibitem{Bai:2003ww}
  J.~Z.~Bai {\it et al.}  [BES Collaboration],
  Phys.\ Rev.\ D {\bf 68}, 052003 (2003)
  [hep-ex/0307058].

\bibitem{Ablikim:2013hq}
  M.~Ablikim {\it et al.}  [BESIII Collaboration],
  Phys.\ Rev.\ D {\bf 87}, no. 9, 092009 (2013)
  [Phys.\ Rev.\ D {\bf 87}, no. 11, 119901 (2013)]
  [arXiv:1301.0053 [hep-ex]].



\bibitem{Dobbs:2015dwa}
  S.~Dobbs, A.~Tomaradze, T.~Xiao and K.~K.~Seth,
  Phys.\ Rev.\ D {\bf 91}, no. 5, 052006 (2015)
  [arXiv:1502.01686 [hep-ex]].



\bibitem{Lepage:1980fj}
  G.~P.~Lepage and S.~J.~Brodsky,
  Phys.\ Rev.\ D {\bf 22}, 2157 (1980).



\bibitem{Efremov:1979qk}
  A.~V.~Efremov and A.~V.~Radyushkin,
  Phys.\ Lett.\ B {\bf 94}, 245 (1980).



\bibitem{Duncan:1979hi}
  A.~Duncan and A.~H.~Mueller,
  Phys.\ Rev.\ D {\bf 21}, 1636 (1980).



\bibitem{Ji:1996nm}
  X.~D.~Ji,
  Phys.\ Rev.\ D {\bf 55}, 7114 (1997)
  [hep-ph/9609381];
  J.\ Phys.\ G {\bf 24}, 1181 (1998)
  [hep-ph/9807358].

\bibitem{Belitsky:1998gc}
  A.~V.~Belitsky and D.~Mueller,
  Nucl.\ Phys.\ B {\bf 537}, 397 (1999)
  [hep-ph/9804379].

\bibitem{Ji:2002aa}
  X.~D.~Ji and F.~Yuan,
  Phys.\ Lett.\ B {\bf 543}, 66 (2002);
M.~Burkardt, X.~D.~Ji and F.~Yuan,
  Phys.\ Lett.\ B {\bf 545}, 345 (2002)
  [hep-ph/0205272].
\bibitem{Fleming:2004rk}
  S.~Fleming and A.~K.~Leibovich,
  Phys.\ Rev.\ D {\bf 70}, 094016 (2004)
  [hep-ph/0407259]; S.~Fleming, C.~Lee and A.~K.~Leibovich,
  Phys.\ Rev.\ D {\bf 71}, 074002 (2005).

\bibitem{Cakir:1994jf}
  M.~B.~Cakir and G.~R.~Farrar, Phys.\ Rev.\ D {\bf 50}, 3268 (1994)[hep-ph/9402203].
  
\bibitem{He:2002hr}
  X.~G.~He, H.~Y.~Jin and J.~P.~Ma,
  Phys.\ Rev.\ D {\bf 66}, 074015 (2002)
  [hep-ph/0203191].
\bibitem{Melis:2004ni}
  M.~Melis, F.~Murgia and J.~Parisi, Phys.\ Rev.\ D {\bf 70}, 034021 (2004)[hep-ph/0404070].
\bibitem{Ma:2001tt}
  J.~P.~Ma,
  Nucl.\ Phys.\ B {\bf 605}, 625 (2001)
  [Nucl.\ Phys.\ B {\bf 611}, 523 (2001)]
  [hep-ph/0103237].



\bibitem{Bodwin:1994jh}
  G.~T.~Bodwin, E.~Braaten and G.~P.~Lepage,
  Phys.\ Rev.\ D {\bf 51}, 1125 (1995)
  [Phys.\ Rev.\ D {\bf 55}, 5853 (1997)]
  [hep-ph/9407339].

\bibitem{Chase:1980hj}
  M.~K.~Chase,
  Nucl.\ Phys.\ B {\bf 174}, 109 (1980).
\bibitem{Belitsky:1998vj}
  A.~V.~Belitsky and D.~Mueller,
  Nucl.\ Phys.\ B {\bf 527}, 207 (1998)
  [hep-ph/9802411].
\bibitem{Ma:2013yla}
  Y.~Q.~Ma, J.~W.~Qiu and H.~Zhang,
  Phys.\ Rev.\ D {\bf 89}, no. 9, 094029 (2014)
  [arXiv:1311.7078 [hep-ph]];Phys.\ Rev.\ D {\bf 89}, no. 9, 094030 (2014)
  [arXiv:1401.0524 [hep-ph]].
\bibitem{Wang:2013ywc}
  X.~P.~Wang and D.~Yang,
  JHEP {\bf 1406}, 121 (2014)
  [arXiv:1401.0122 [hep-ph]].

\bibitem{Kroll:2002nt}
  P.~Kroll and K.~Passek Kumericki,Phys.\ Rev.\ D {\bf 67}, 054017 (2003)[hep-ph/0210045].
  
\bibitem{Close:2005vf}
  F.~E.~Close and Q.~Zhao,
  Phys.\ Rev.\ D {\bf 71}, 094022 (2005)
  [hep-ph/0504043].
\bibitem{Cho:2015rsa}
  Y.~M.~Cho, X.~Y.~Pham, P.~Zhang, J.~J.~Xie and L.~P.~Zou,
  Phys.\ Rev.\ D {\bf 91}, no. 11, 114020 (2015)[arXiv:1503.08890 [hep-ph]].
\bibitem{Gui:2012gx}
  L.~C.~Gui, Y.~Chen, G.~Li, C.~Liu, Y.~B.~Liu, J.~P.~Ma, Y.~B.~Yang and J.~B.~Zhang,
  Phys.\ Rev.\ Lett.\  {\bf 110}, 021601 (2013)
  [arXiv:1206.0125 [hep-lat]].
\bibitem{Agashe:2014kda}
  K.~A.~Olive {\it et al.}  [Particle Data Group Collaboration],
  Chin.\ Phys.\ C {\bf 38}, 090001 (2014).
\bibitem{Zhu:2015jha}
  R.~Zhu,
  arXiv:1507.02031 [hep-ph]; C.~F.~Qiao, P.~Sun, D.~Yang and R.~L.~Zhu,
  Phys.\ Rev.\ D {\bf 89}, no. 3, 034008 (2014)
  [arXiv:1209.5859 [hep-ph]].
\bibitem{Qiao:2014pfa}
  C.~F.~Qiao and R.~L.~Zhu,
  Phys.\ Rev.\ D {\bf 89}, 074006 (2014)
  [arXiv:1403.1918 [hep-ph]];
  C.~F.~Qiao and R.~L.~Zhu,
  Phys.\ Rev.\ D {\bf 87}, 014009 (2013)
  [arXiv:1208.5916 [hep-ph]].
\bibitem{Wang1}
   W.~Wang and R.~L.~Zhu,arXiv:1501.04493 [hep-ph].
\bibitem{Cheng:2006hu}
  H.~Y.~Cheng, C.~K.~Chua and K.~F.~Liu,
  Phys.\ Rev.\ D {\bf 74}, 094005 (2006)
  [hep-ph/0607206].
\bibitem{Lee:1999kv}
  W.~J.~Lee and D.~Weingarten,
  Phys.\ Rev.\ D {\bf 61}, 014015 (2000)
  [hep-lat/9910008].


\bibitem{Passarino:1978jh}
  G.~Passarino and M.~J.~G.~Veltman,
  Nucl.\ Phys.\ B {\bf 160}, 151 (1979).
\bibitem{Hahn:1998yk}
  T.~Hahn and M.~Perez-Victoria,
  Comput.\ Phys.\ Commun.\  {\bf 118}, 153 (1999)
  [hep-ph/9807565].
\end{thebibliography}
\end{document}